\documentclass[aps,pre,preprint]{revtex4}
\usepackage{amsmath}
\usepackage{amssymb}
\usepackage{epsfig}
\usepackage{graphicx}
\usepackage[T1]{fontenc}
\usepackage[utf8]{inputenc}
\usepackage{color}
\usepackage[normalem]{ulem}
\usepackage{tikz}
\usetikzlibrary{positioning}

\begin{document}

\title{Transient behavior of damage spreading in the two-dimensional Blume-Capel ferromagnet}

\author{Ajanta Bhowal Acharyya}

\affiliation{Department of Physics, Lady Brabourne College, Kolkata-700017, India}

\author{Muktish Acharyya}

\affiliation{Department of Physics, Presidency University,
	86/1 College Street, Kolkata-700073, India}

\author{Erol Vatansever}

\affiliation{Department of Physics, Dokuz Eyl\"{u}l University, TR-35160, Izmir, Turkey}

\affiliation{Centre for Fluid and Complex Systems, Coventry
	University, Coventry, CV1 5FB, United Kingdom}

\author{Nikolaos G. Fytas}

\affiliation{Centre for Fluid and Complex Systems, Coventry
	University, Coventry, CV1 5FB, United Kingdom}

\affiliation{Institut für Physik, Technische Universität Chemnitz, 09107 Chemnitz, Germany}

\date{\today}

\begin{abstract}
We study the transient behavior of damage propagation in the two-dimensional spin-$1$ Blume-Capel model using Monte Carlo simulations with Metropolis dynamics. We find that, for a particular region in the second-order transition regime of the crystal field--temperature phase diagram of the model, the average Hamming distance decreases exponentially with time in the weakly damaged system. Additionally, its rate of decay appears to depend linearly on a number of Hamiltonian parameters, namely the crystal field, temperature, applied magnetic field, but also on the amount of damage. Finally, a comparative study using Metropolis and Glauber dynamics indicates a slower decay rate of the 
average Hamming distance for the Glauber protocol.
\end{abstract}

\maketitle




\section{Introduction}
\label{sec:1}

A system is said to exhibit damage spreading (DS) if the \emph{distance} between two of its replicas that evolve under the same thermal noise, but from slightly different initial conditions, increases with time. DS was first introduced by Kauffman for examining the metabolic stability and epigenesis in randomly constructed genetic nets~\cite{Kauffman:69} and now it has evolved into an important phenomenon in physics. It is used in equilibrium~\cite{Grassberger:95} for measuring accurately dynamic exponents and also out of equilibrium, to study the influence of initial conditions on the temporal evolution of various systems. From a theoretical point of view, many studies engaged in unraveling the DS in systems of many interacting particles. Typical examples include Ising~\cite{Stanley, Coniglio, Caer, Cruz, Arcangelis1, Arcangelis2, Barber, Mariz1, Mariz2, Moreira, Nobre, Tamarit, Wang, Hinrichsen, Svenson, Puzzo1}, Potts~\cite{Bibiano, Silva, Luz, Redinz, Anjos, Mariz3}, Ashkin-Teller~\cite{Mariz3, Anjos2, Anjos3}, XY~\cite{Golinelli, Chiu}, Bak-Sneppen~\cite{Tirnakli}, sandpile~\cite{Ajanta}, opinion dynamics~\cite{Khaleque}, and Heisenberg~\cite{Miranda, Rojdestvenski, Costa1, Costa2} models.

How a perturbation spreads throughout a spin system strongly depends on its parameters, such as the temperature, initial damage, etc. It is well-established that in this process more than one regime is expected even for the simplest models. For example, depending on the temperature $T$, typical ferromagnets can either be in the chaotic (the two configurations remain apart from each other) or the frozen (the initial damage heals leading to the reunification of the two different configurations) regime. There is a special temperature, the so-called spreading temperature, $T_{\rm s}$, that separates these two different regimes. By performing Monte-Carlo simulations and using various spin dynamics, such as Metropolis~\cite{Mariz2}, Glauber~\cite{Stanley, Mariz2}, and heat-bath~\cite{Mariz2}, it has been shown that $T_{\rm s} \equiv T_{\rm c}$ for the two dimensional Ising ferromagnet, the well-known Curie temperature. DS in the kinetic Ising model has been investigated extensively by Vojta~\cite{Vojta} in an effective field approach where the regular and chaotic phases were identified by the fixed point analysis of the master equation. We refer the reader to Ref.~\cite{Puzzo2} for an extensive and detailed review on DS phenomena. 

Most of the relevant work done so far refers to the spin-$1/2$ Ising model. To the best of our knowledge, there is a limited number of studies regarding DS properties of spin-$S$ Blume-Capel models for $S \geq 1$~\cite{Liu1, Liu2, Costa3, Jun}. In Ref.~\cite{Liu1} DS was used to investigate the general integer and half-integer spin-$S$ Blume-Capel model on the square lattice within a Metropolis-type dynamics. These authors studied how the temperature ($T$) and crystal field ($\Delta$) influence the average Hamming distance, reporting on the existence of a multi-critical point along the order-disorder transition line for the $S = 1$ and $2$ cases. This multi-critical behavior does not seem to appear in the $\Delta - T$ plane for the $S = 3/2$ and $5/2$ cases. A similar analysis was performed for the mixed-spin Ising model consisting of spin-$1/2$ and spin-$1$
states under the presence of a quenched crystal field, again via Monte-Carlo simulations~\cite{Liu2}. The numerical results obtained suggested that there might be a dynamical tricritical point along the phase boundary. Although the main conclusion emerging from these works is that both $T$ and $\Delta$ play a key role in DS, still several important questions related to transient phenomena remain. In this context, the first port of call is a clarification of the functional form that describes the time evolution of the average Hamming distance and its dependence on the Hamiltonian parameters and the amount of damage in the system.

In the present work we scrutinize the transient behavior of DS using as a platform the spin-$1$ Blume-Capel model. We elaborate on how a small perturbation in the initial configuration of the lattice decays, using extensive Monte-Carlo simulations with Metropolis dynamics. The main outcome of our analysis is that, for a particular regime in the $\Delta - T$ plane, the average Hamming distance decreases exponentially with time in the weakly damaged system and its rate of decay depends linearly on several Hamiltonian parameters, but also on the amount of damage in the system. Some additional complementary simulations using Glauber dynamics are performed for comparative reasons.

The remainder of this paper is organized as follows: In Sec.~\ref{sec:2} we introduce the model and outline our numerical protocol and in Sec.~\ref{sec:3} we present the analysis of numerical data. The paper concludes with a summary and an outlook for future research in Sec.~\ref{sec:4}.

\section{Model and simulation scheme}
\label{sec:2}

The Hamiltonian of the Blume-Capel ferromagnet in the presence of an external magnetic field $h$ reads as~\cite{Blume,Capel}
\begin{equation}
\label{eq:hamiltonian}
\mathcal{H} = -J\sum_{\langle ij \rangle}S_iS_j + \Delta\sum_i S_i^2 - h\sum_{i} S_i,
\end{equation} 
where the spin variables take on the values $S_i\in\{-1,0,+1\}$ and $\langle ij \rangle$ indicates summation over nearest
neighbors. The crystal
field (also called single-ion anisotropy) $\Delta$ controls the density of vacancies, \emph{i.e.}, sites with
$S_i = 0$. When $\Delta \rightarrow -\infty$ vacancies are suppressed and the model maps onto the spin-1/2 Ising model. The Blume-Capel model has been studied extensively (for a review see Ref.~\cite{zierenberg17}) and the phase diagram in the $\Delta -T$ plane is known to a high accuracy: For small $\Delta$ there is a line of continuous transitions
between the ferromagnetic and paramagnetic phases that crosses the $\Delta = 0$ axis
at $T_{0}\approx 1.693$~\cite{malakis10}. For large $\Delta$, on the other hand, the
transition becomes discontinuous and it meets the $T=0$ line at
$\Delta_0 = zJ/2$~\cite{Capel}, where $z = 4$ is the coordination number (here we
set $J = 1$, and also $k_{\rm B} = 1$, to fix the temperature scale). The two line
segments meet in a tricritical point estimated to be at
$(\Delta_{\rm t}\approx1.966,T_{\rm t}\approx0.608)$~\cite{kwak2015,jung17}. It is
well established that the second-order transition regime, where we also focus on this work, belongs to the universality class of the two-dimensional Ising model~\cite{zierenberg17}.

We consider the Blume-Capel model~(\ref{eq:hamiltonian}) on the square lattice where all the sites have spin values $S_i = +1$. From this original lattice a replica 
lattice is generated, where a fraction ($p$) of sites differs from the original one. These damaged sites of 
the replica lattice are chosen randomly and are set to $S_i=-1$. The two lattices are then allowed to evolve in time using a random updating rule and periodic boundary conditions at all directions. We observe how the number of sites having different spin values in the two lattices, defined through the parameter~\cite{Jun, Puzzo2}
\begin{equation}
\label{eq:delta}
D = \frac{1}{L^{2}}{\sum_{i=1}^{L^{2}} (1- \delta_{S_i^{\rm o}, S_i^{\rm r}})},
\end{equation} 
changes with time. In Eq.~(\ref{eq:delta}) above, $L$ denotes the linear size of the system, $\delta$ is the Kronecker delta, $S_i^{\rm o}$ the value of the spin component  at the $i$th site of original lattice and $S_i^{\rm r}$ the value of the spin component at the $i$th site of the 
replica lattice. From the above discussion, it is clear that there are six possibilities for a chosen site $i$ having a damage at a given time $t$ between $S_i^{\rm o}$ and $S_i^{\rm r}$: $(+1, 0)$, $(+1, -1)$, $(0, +1)$, $(0, -1)$, $(-1, +1)$, and $(-1, 0)$. 

\begin{figure}
	\centering
	\includegraphics[width=8 cm]{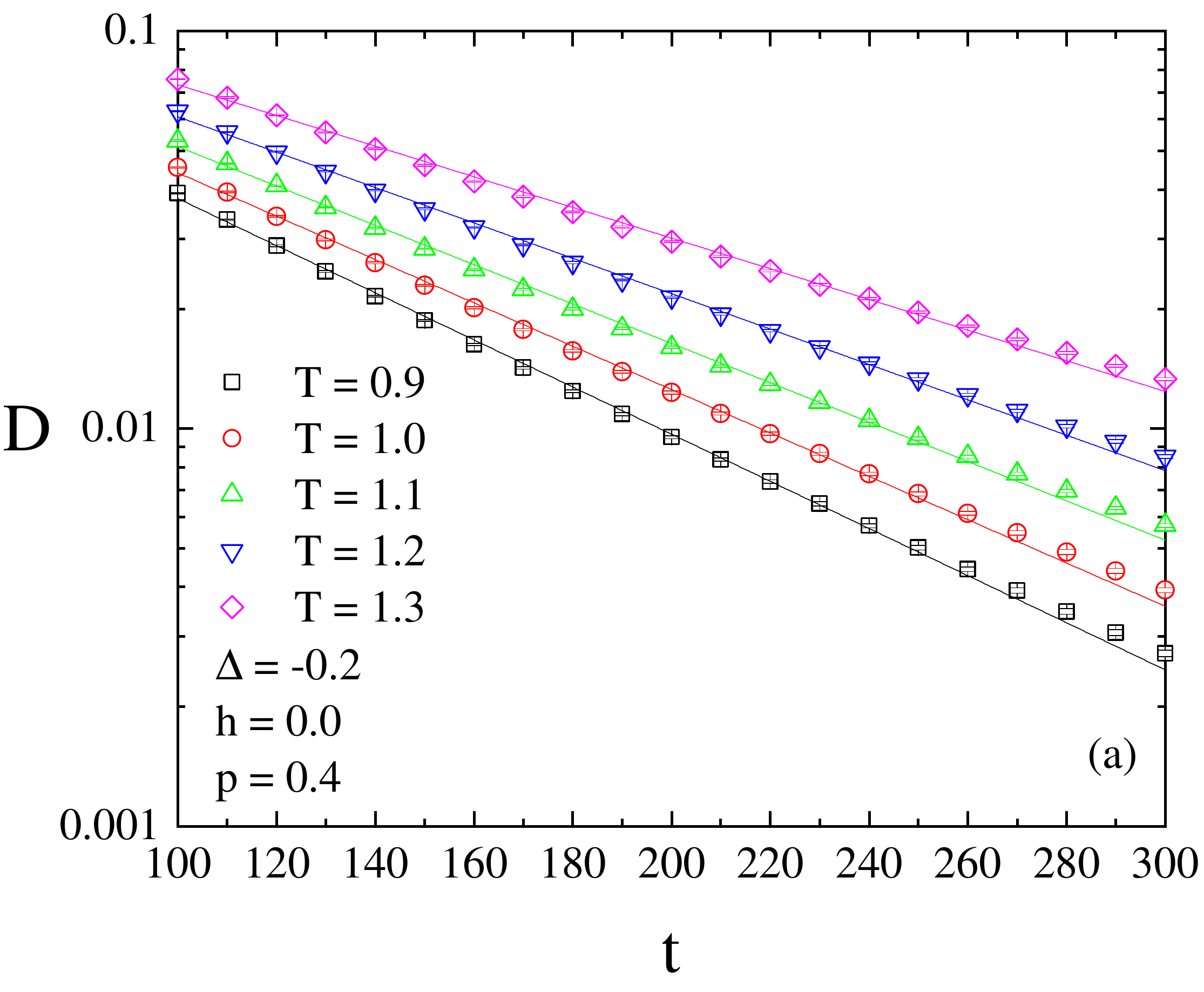}\\
	\includegraphics[width=8 cm]{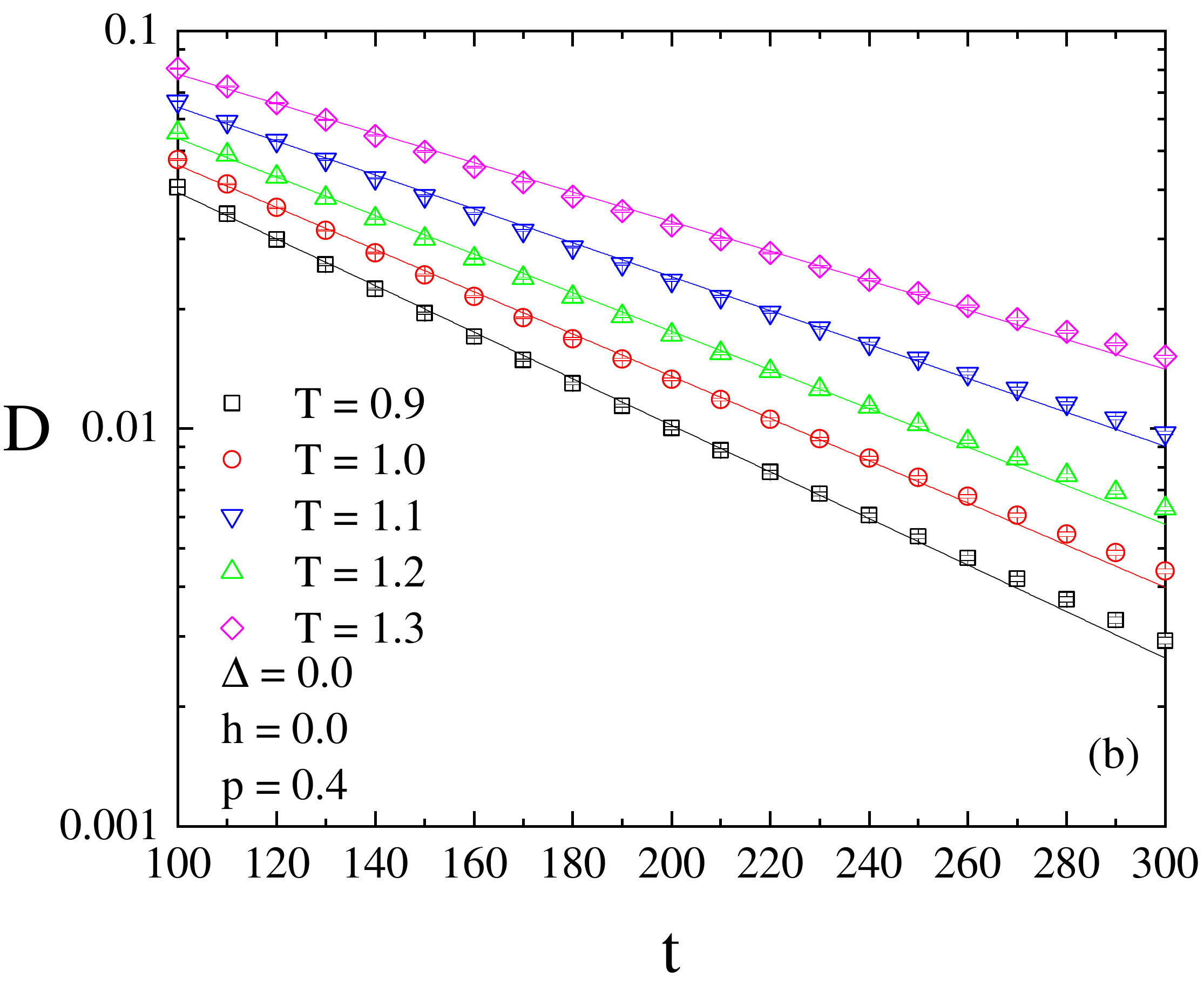}\\
	\includegraphics[width=8 cm]{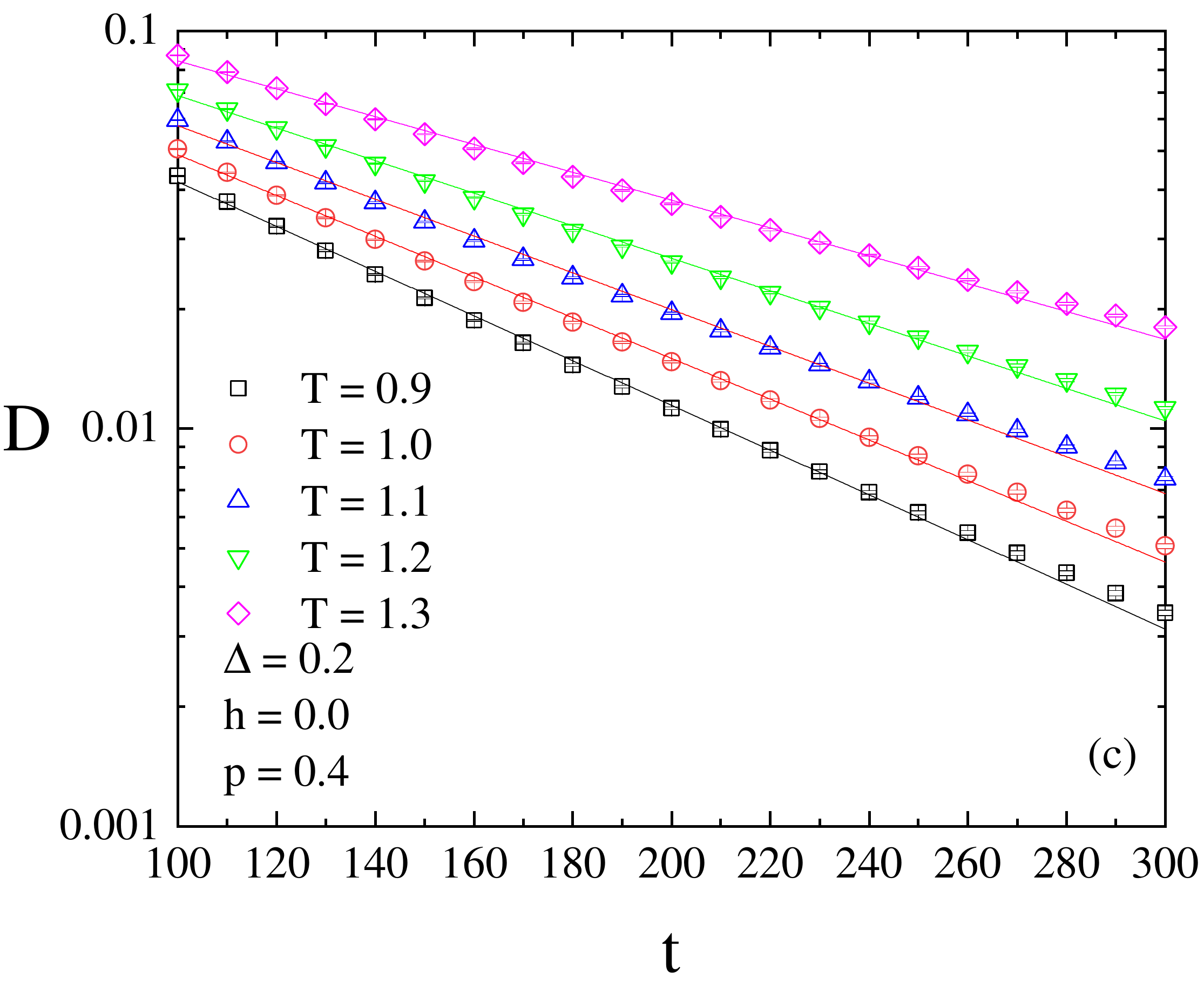}
	\caption{\label{Fig1} $D(t)$ curves for a set of selected temperatures $T$ and three values of the crystal field as specified in the panels: $\Delta = -0.2$ (a), $\Delta = 0$ (b), and $\Delta = 0.2$ (c). $h = 0$ and $p = 0.4$ in all panels.}
\end{figure}

In order to generate a spin configuration $C(t)=\{S_i(t)\}$, where $i = 1, 2, 3, \ldots, L^2$, we follow the steps given below:
During each time interval $\delta(t) = 1/L^2$, a lattice site $i$ is randomly selected among the $L^2$ options. Then, a spin value $C(t+\delta(t))$ is proposed via
\begin{equation}
\label{eq:cases1}
C\left(t+\delta t\right)=
\begin{cases}
1,       & \text{if} \; 0 \leq r_{i_1} < 1/3  \\
0,        & \text{if} \; 1/3 \leq r_{i_1} < 2/3 \\
-1,        & \text{if} \; 2/3 \leq r_{i_1} < 1,
\end{cases}
\end{equation}
where $r_{i_1}$ is a random number generated at time $t$, such that $0\leq r_{i_1} \leq 1$. Accordingly, the selected spin updates following the well-known Metropolis dynamics rule~\cite{landau_book}
\begin{equation}
S_i\left(t+\delta t\right)=
\begin{cases}
C\left(t+\delta t\right),       & \text{if}\; P_i(t) \geq r_{i_2} \\
S_i(t),        &  \text{if}\; P_i(t) $<$ r_{i_2},
\end{cases}
\end{equation}
where $r_{i_2}$ is another random number, and $P_i(t)=\exp(-\Delta \mathcal{H}/T)$. Here, $\Delta \mathcal{H}$ is the energy change originating from the spin-flip operation. 

\begin{figure}
	\centering
    \includegraphics[width=8 cm]{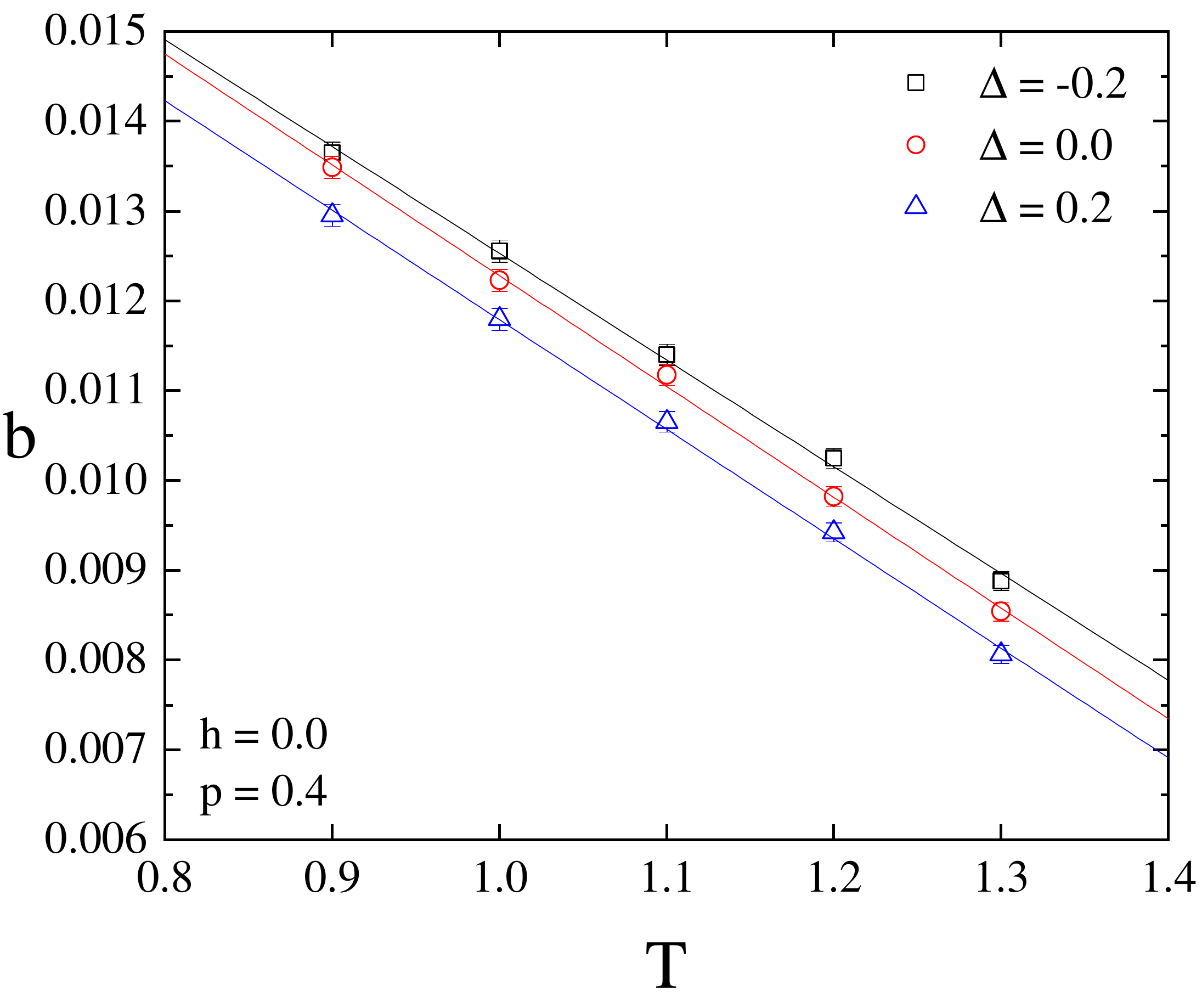}
	\caption{Decay constant $b$, see Eq.~(\ref{eq:Dexp}), versus temperature $T$ for the same set of parameters as in Fig.~\ref{Fig1}.}  
    \label{Fig2}
\end{figure}

Using this numerical protocol we simulated the Blume-Capel model~(\ref{eq:hamiltonian}) for a system with linear size $L = 1024$ and for a range of the Hamiltonian parameters $\{\Delta, T,  h\}$, but also while varying the fractional amount of damage $p$. We also undertook a comparative study of DS under different dynamics rules, namely Metropolis versus Glauber, for some particular values of the control parameters. In our analysis, and in order to increase statistical accuracy, $D$ was computed by taking the average over $200$ different random configurations of the replica lattice. Errors were estimated via the jackknife method and in most cases appeared to be smaller than the symbol sizes. We also employed the standard $\chi^2$ test for goodness of the fit. Specifically, we considered a fit as being fair only if $10\% < Q < 90\%$, where $Q$ is the quality-of-fit parameter~\cite{press}.

\begin{figure}
	\centering
	\includegraphics[width=8 cm]{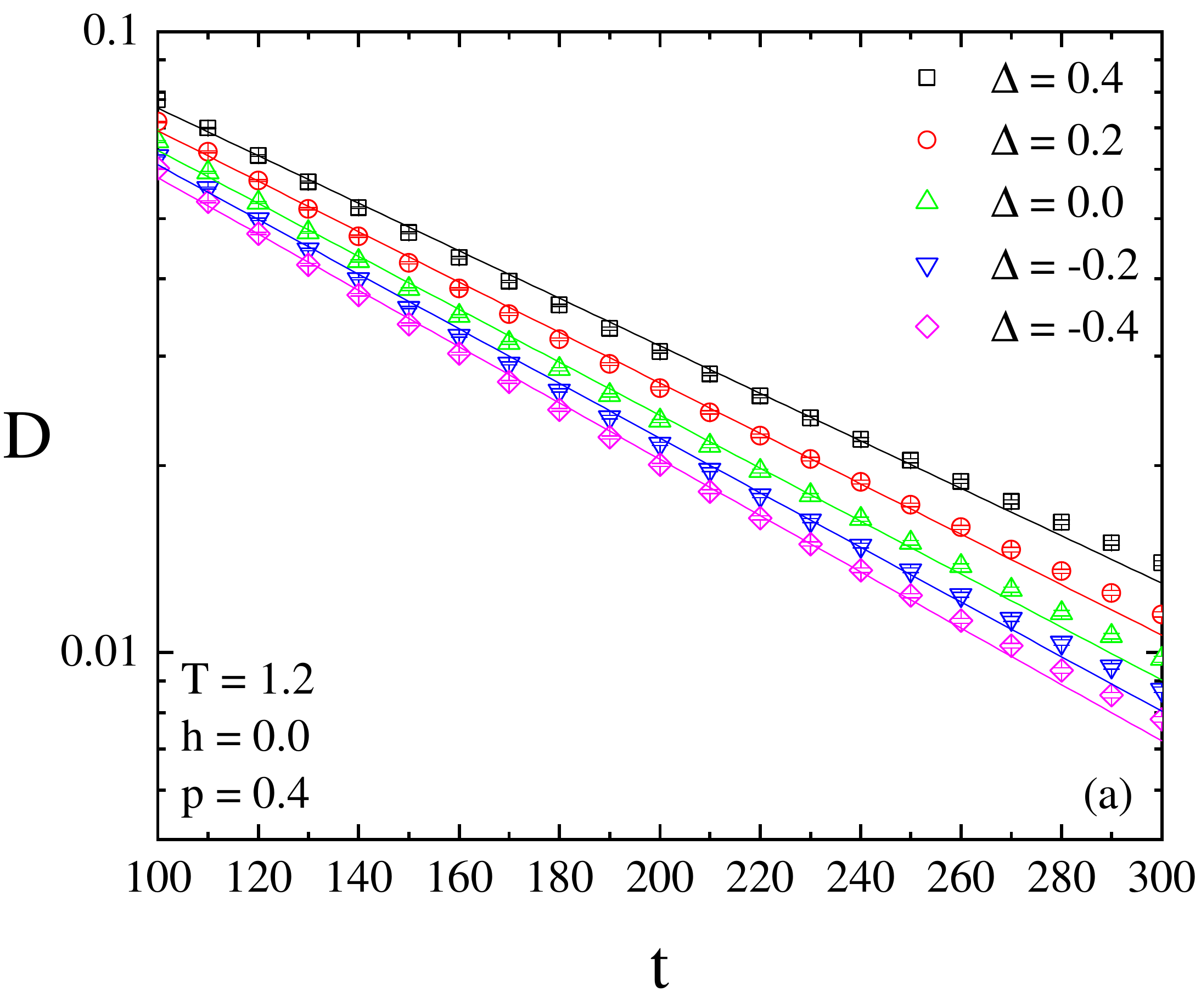}\\
	\includegraphics[width=8 cm]{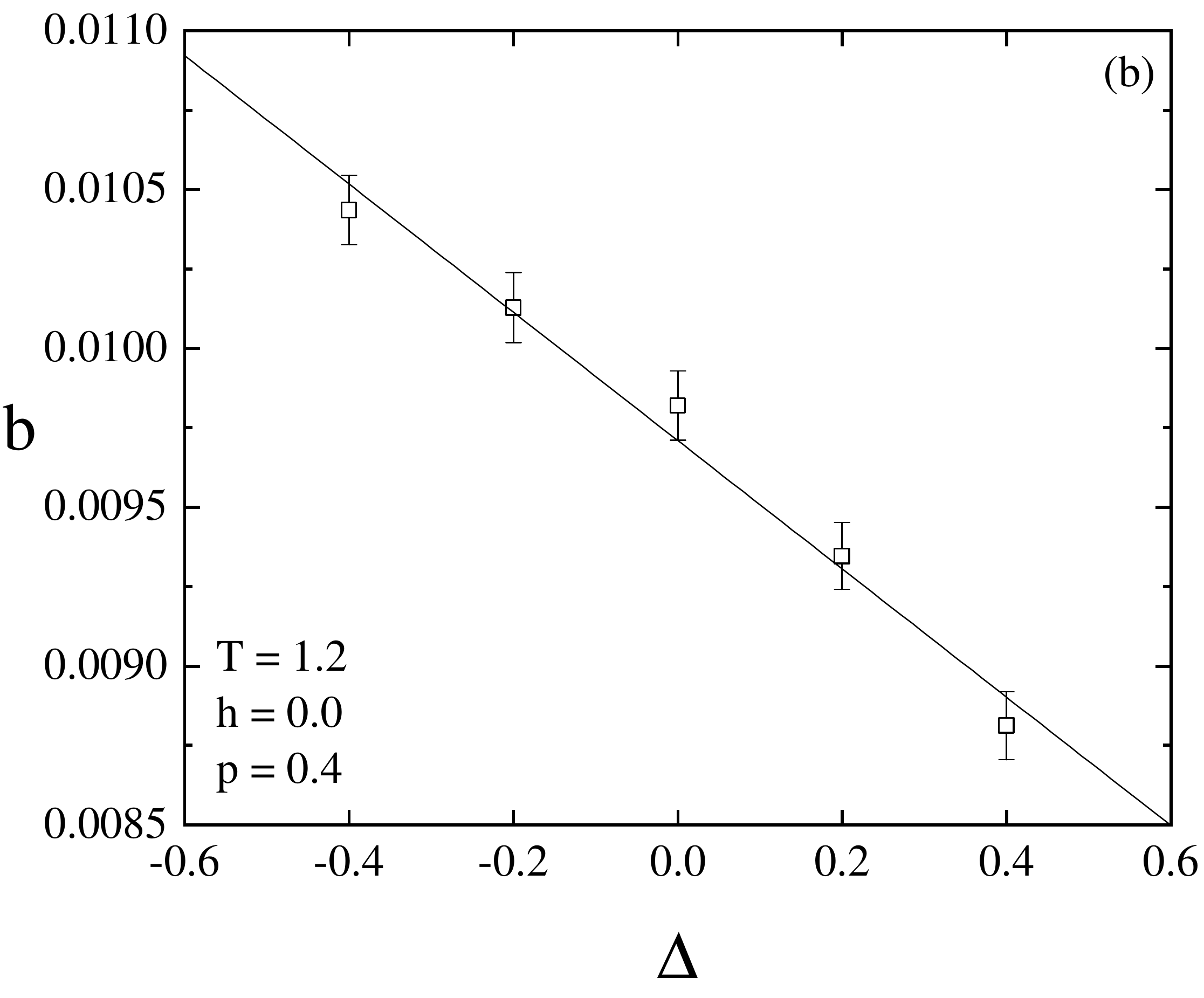}
	\caption{\label{Fig3} (a) $D(t)$ curves for a set of selected crystal fields at fixed values of $T = 1.2$, $h = 0$, and $p = 0.4$. (b) Decay constant $b$ versus crystal field $\Delta$ for the set of parameters outlined in panel (a).}
\end{figure}


\begin{figure}
	\centering
	\includegraphics[width=8 cm]{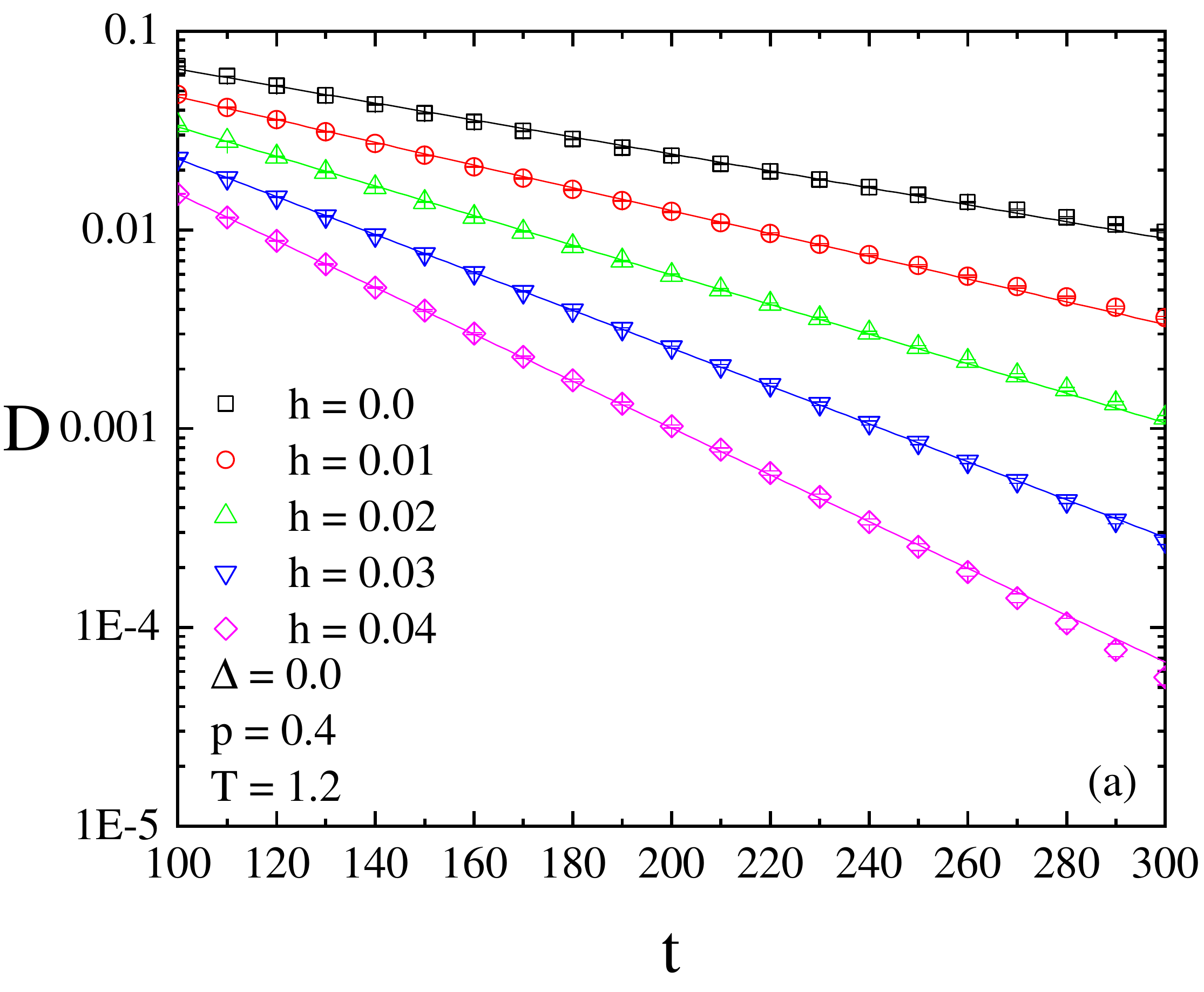}\\
	\includegraphics[width=8 cm]{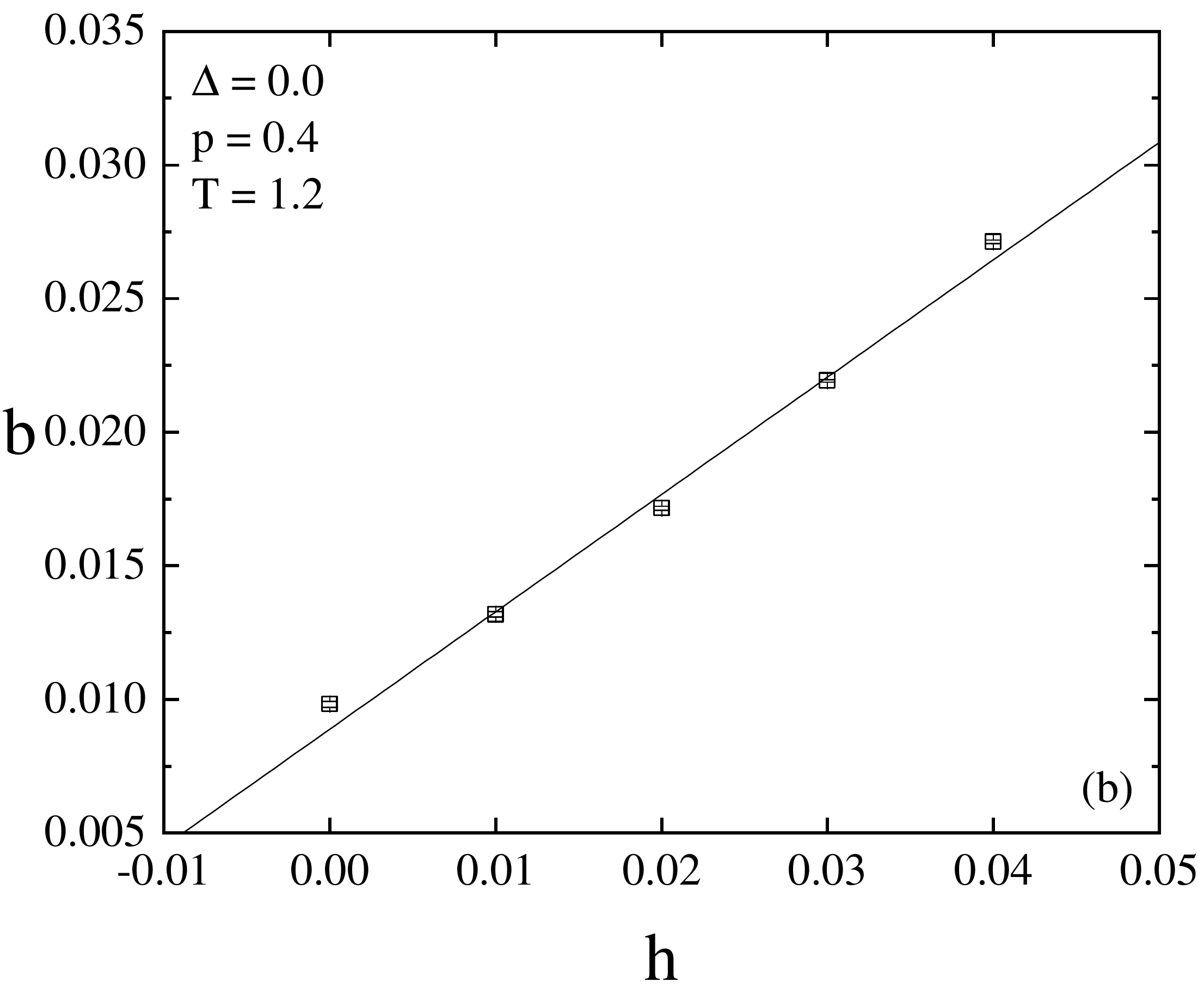}
	\caption{\label{Fig4} (a) $D(t)$ curves for a set of selected magnetic fields at fixed values of $T = 1.2$, $\Delta = 0$, and $p = 0.4$. (b) Decay constant $b$ versus magnetic field $h$ for the set of parameters outlined in panel (a).}
\end{figure}

\section{Results and discussion}
\label{sec:3}

We start the presentation of our results with Fig.~\ref{Fig1} which illustrates the time progress of the DS parameter $D$ in a semi-logarithmic plot. Results for five different values of the temperature are shown in the range $T = \{0.9 - 1.3\}$ and in steps of $0.1$. All curves are obtained at $h = 0$ and $p = 0.4$ and for three selected values of the crystal field. The linear nature of the semi-logarithmic plots indicates clearly that $D$ decreases exponentially with time following a law of the form 
\begin{equation}
\label{eq:Dexp}
D \sim \exp{\left(-bt\right)},
\end{equation}
thus suggesting that damage is healing (instead of spreading) in time. The values of the decay constant $b$ have been estimated by simple linear fits, as shown by the solid lines in all panels, and depend on the chosen temperature. In particular, the temperature variation of the decay constant is depicted in Fig.~\ref{Fig2} for the same three values of $\Delta$ deducing that the magnitude of $b$ decreases linearly with the temperature, within a limited regime away from the phase boundary and sufficiently above $T = 0$. In these two limiting cases ($T \rightarrow T_{\rm c}$ and $T \rightarrow 0$) a different behavior may be expected~\cite{Hinrichsen,Vojta,Grassberger2} -- see also the discussion in Sec.~\ref{sec:4}. Assuming the simple linear functional form $b = sT + \rm const$, as shown by the solid lines in Fig.~\ref{Fig2}, and based on the fitting results we may argue that the slope $\rm s$ is practically independent of $\Delta$ in this regime. We quote the values $s(\Delta = -0.2) = -0.0119(29)$, $s(\Delta = 0) = -0.0123(26)$, and $s(\Delta = 0.2) = -0.0122(26)$.

\begin{figure}
	\centering
	\includegraphics[width=8 cm]{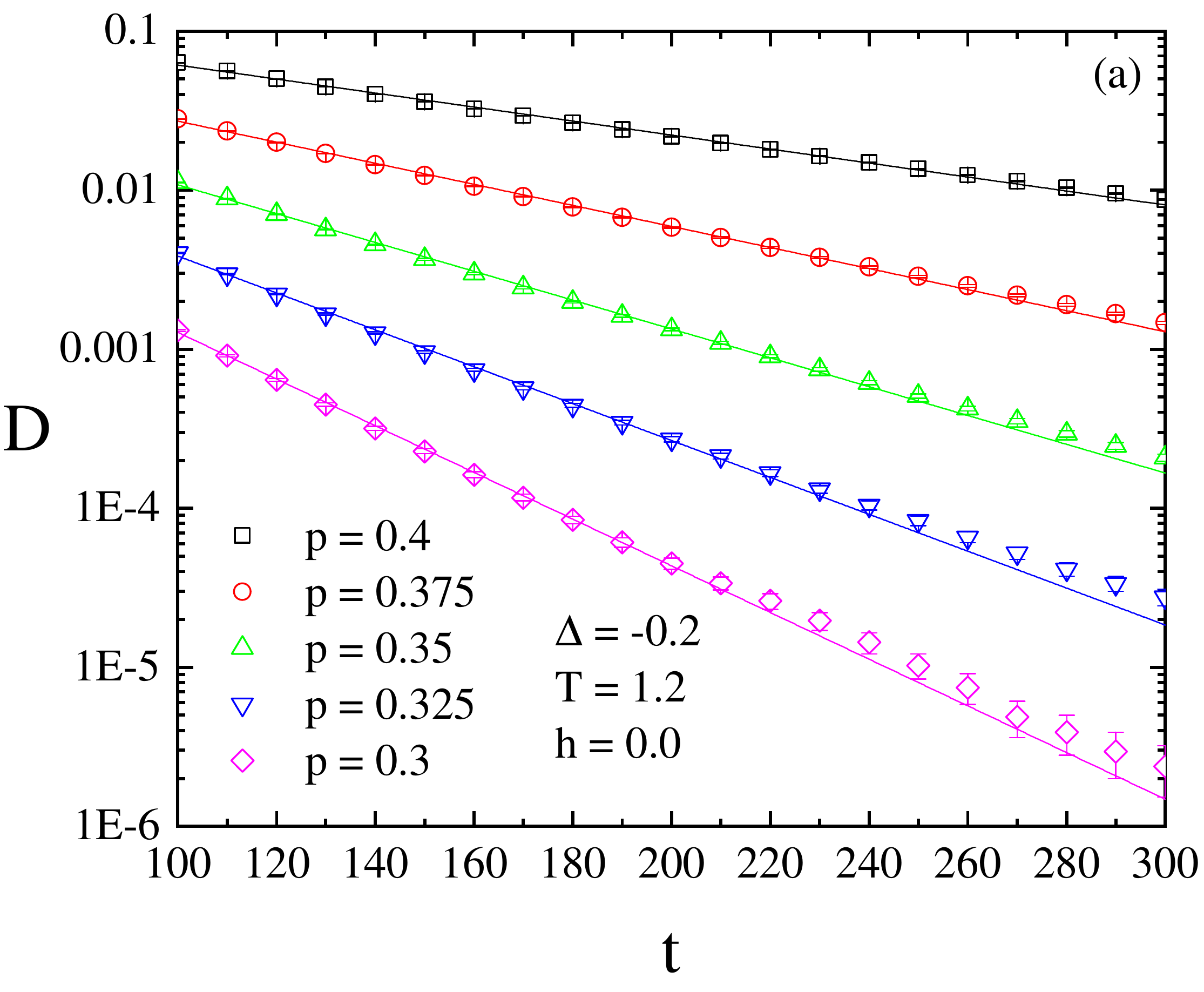}\\
	\includegraphics[width=8 cm]{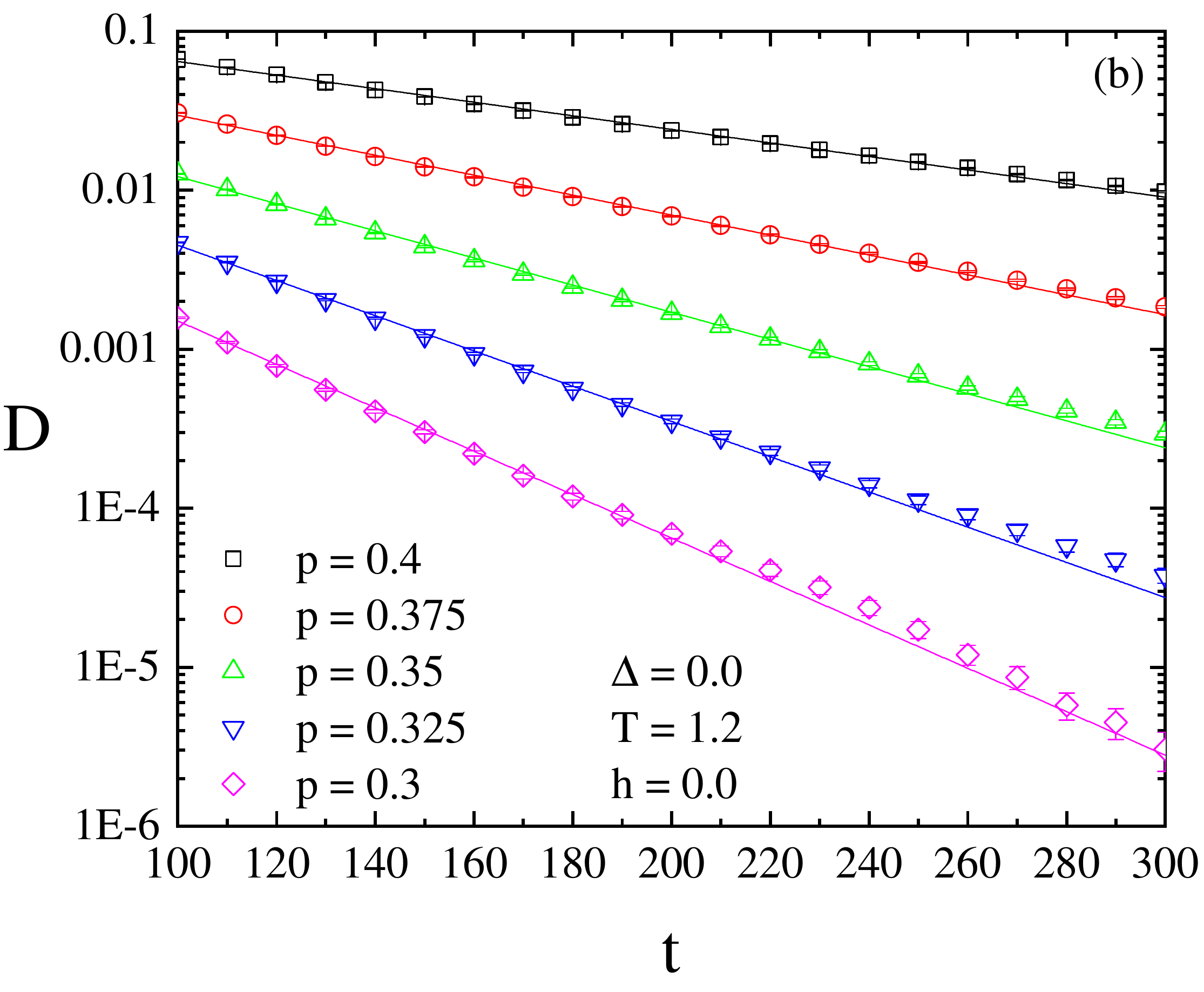}\\
	\includegraphics[width=8 cm]{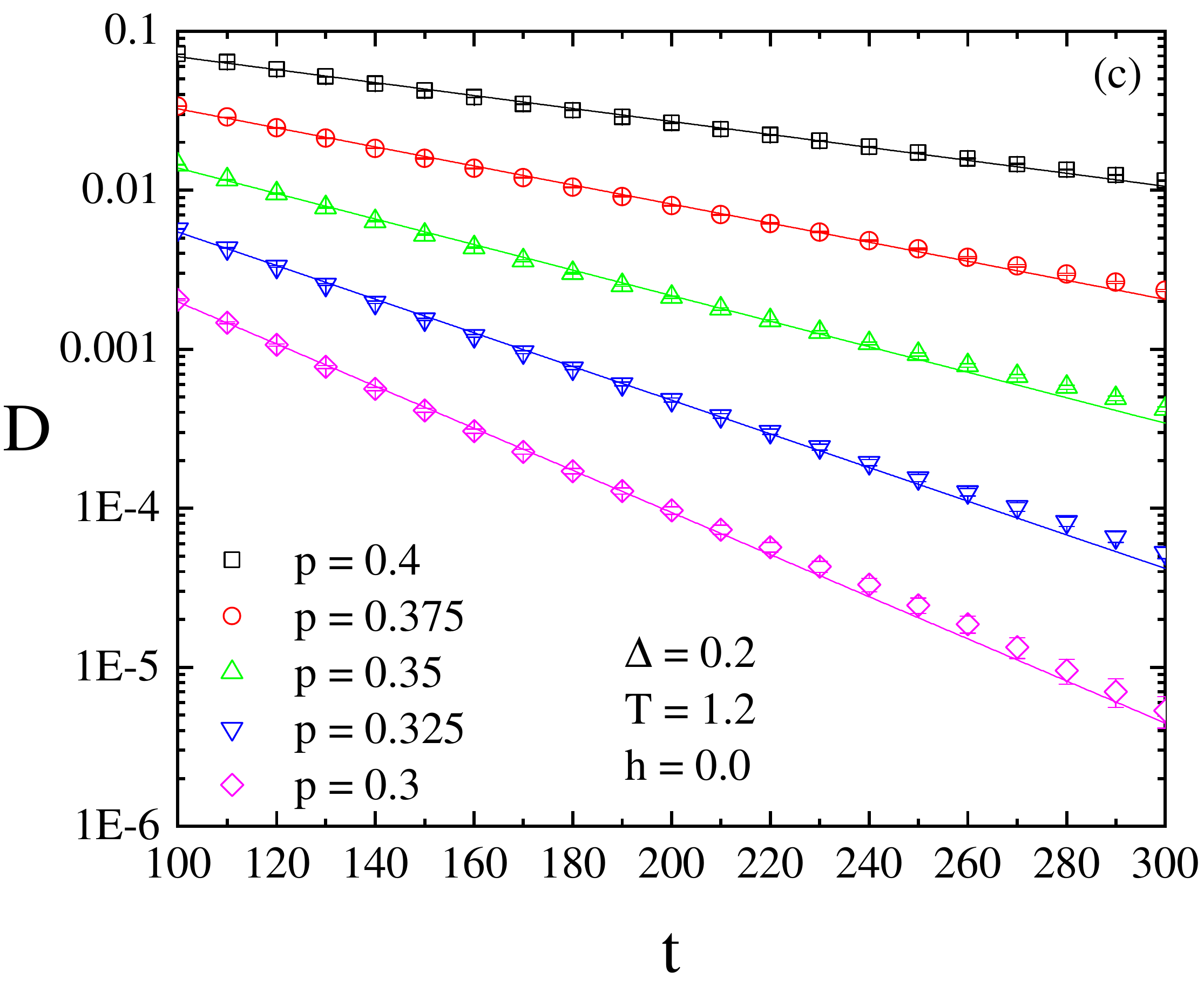}
	\caption{\label{Fig5} $D(t)$ curves for a set of selected $p$-values in the weak-damage regime and three values of the crystal field: $\Delta = -0.2$ (a), $\Delta = 0$ (b), and $\Delta = 0.2$ (c). $T = 1.2$ and $h = 0$ in all panels.}
\end{figure}

\begin{figure}
	\centering
	\includegraphics[width=8 cm]{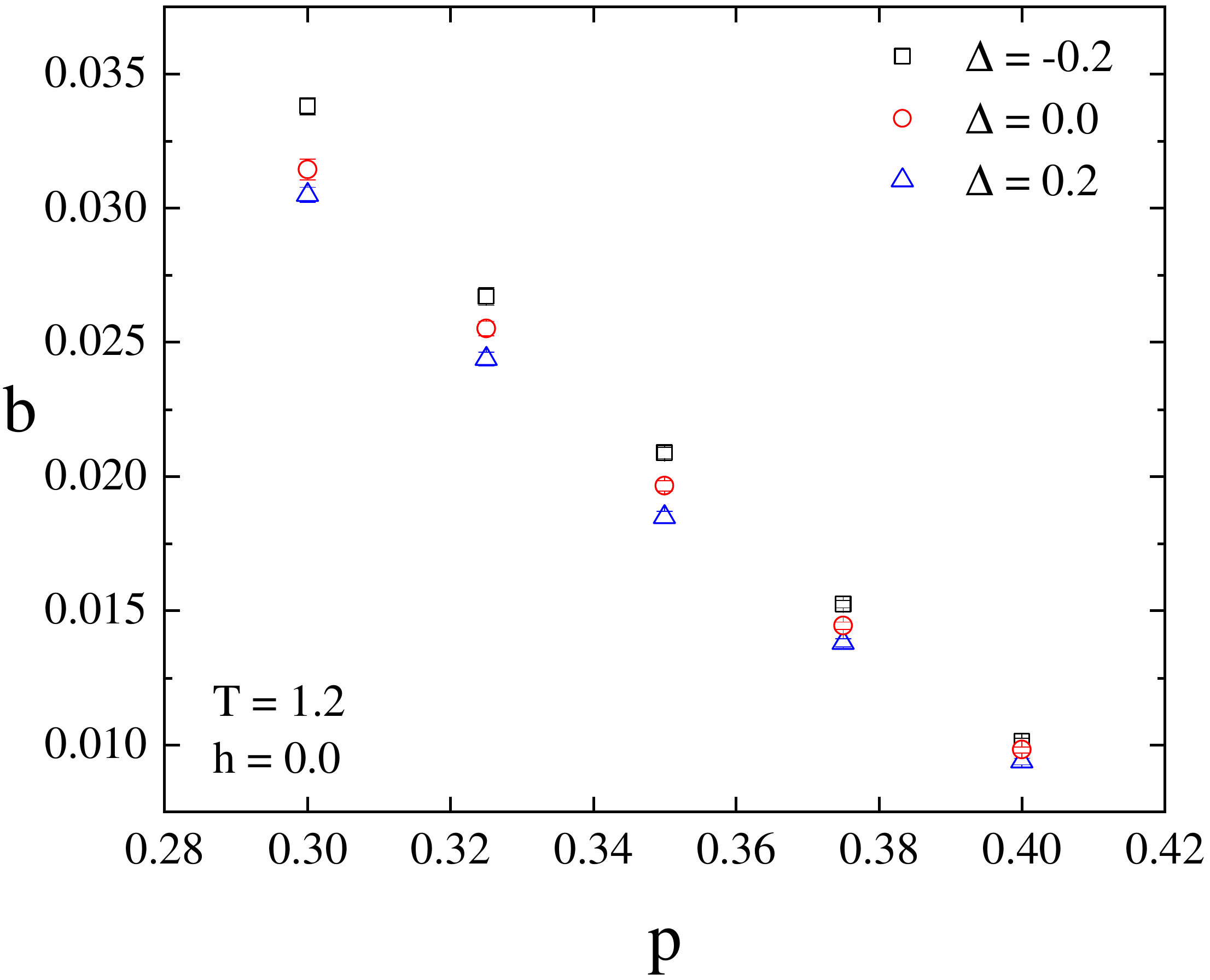}
	\caption{Variation of the decay constant $b$ versus the amount of damage $p$ for three selected values of the crystal field at fixed $T = 1.2$ and $h = 0$.}  
		\label{Fig6}
\end{figure}

Analogously to Fig.~\ref{Fig1}, we also studied the time growth of $D$ at fixed values of $T = 1.2$, $h = 0$, and $p = 0.4$, while varying the crystal field $\Delta$. In Fig.~\ref{Fig3}(a) the set of results for five different values of the crystal field in the range $\Delta = \{-0.4 - 0.4\}$ is shown, again in semi-logarithmic scale, where a similar behavior to Eq.~(\ref{eq:Dexp}) is observed. We know that as the crystal field increases to values $\Delta \geq 0$, $0$-state spins become gradually dominant because these are the configurations that lower the energy. As a result, and in accordance with the numerical data, $D$ tends to get larger as $\Delta$ increases beyond zero, since all spins in the original lattice were initially set to the spin-state $+1$. Now the dependence of the decay constant $b$ on $\Delta$ is presented in panel (b) of Fig.~\ref{Fig3}, where the solid line is the usual linear fit, $b = s\Delta + \rm const$, with a negative slope $s = -0.0020(15)$. 

\begin{figure}
	\centering
	\includegraphics[width=8 cm]{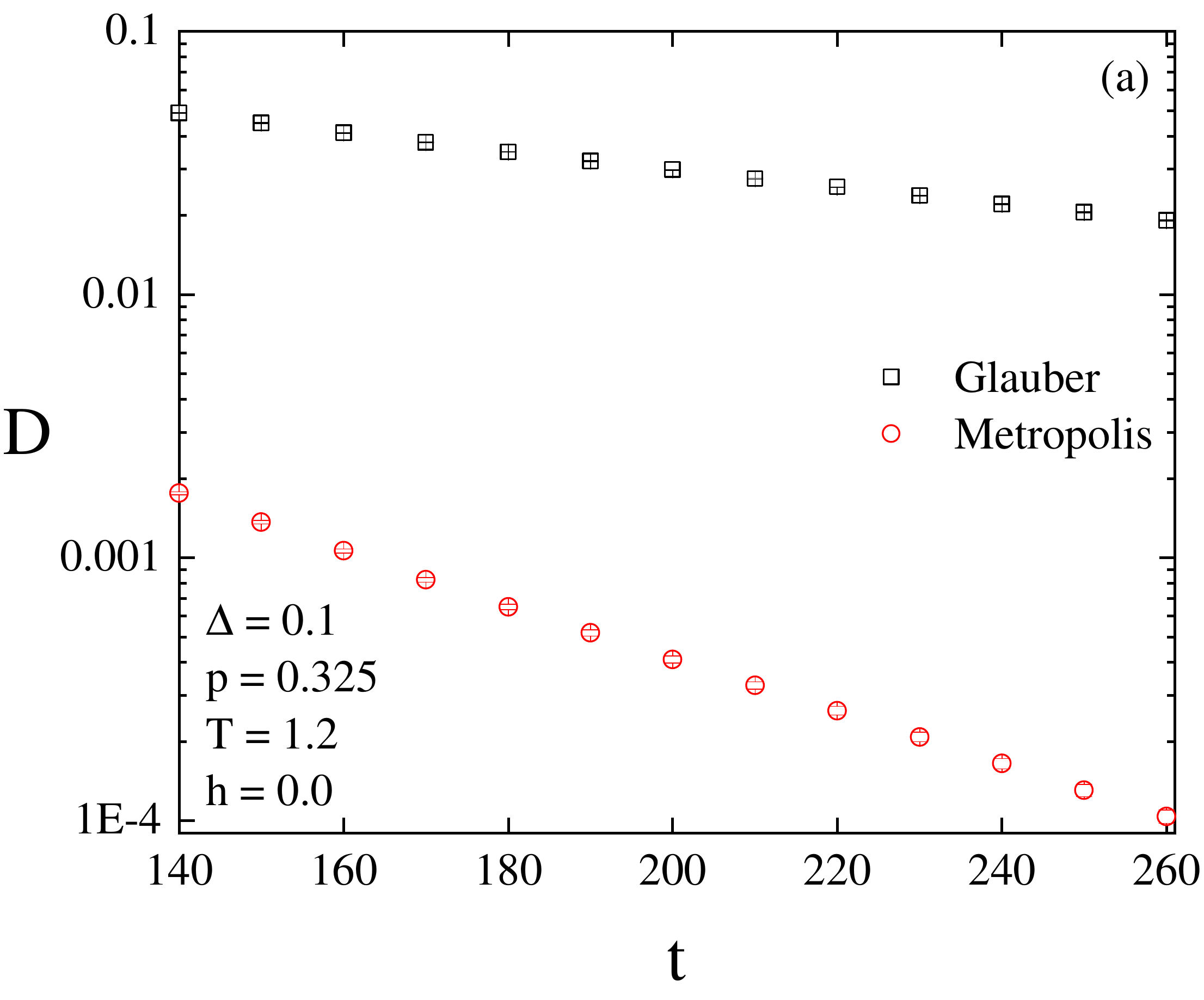}\\
	\includegraphics[width=8 cm]{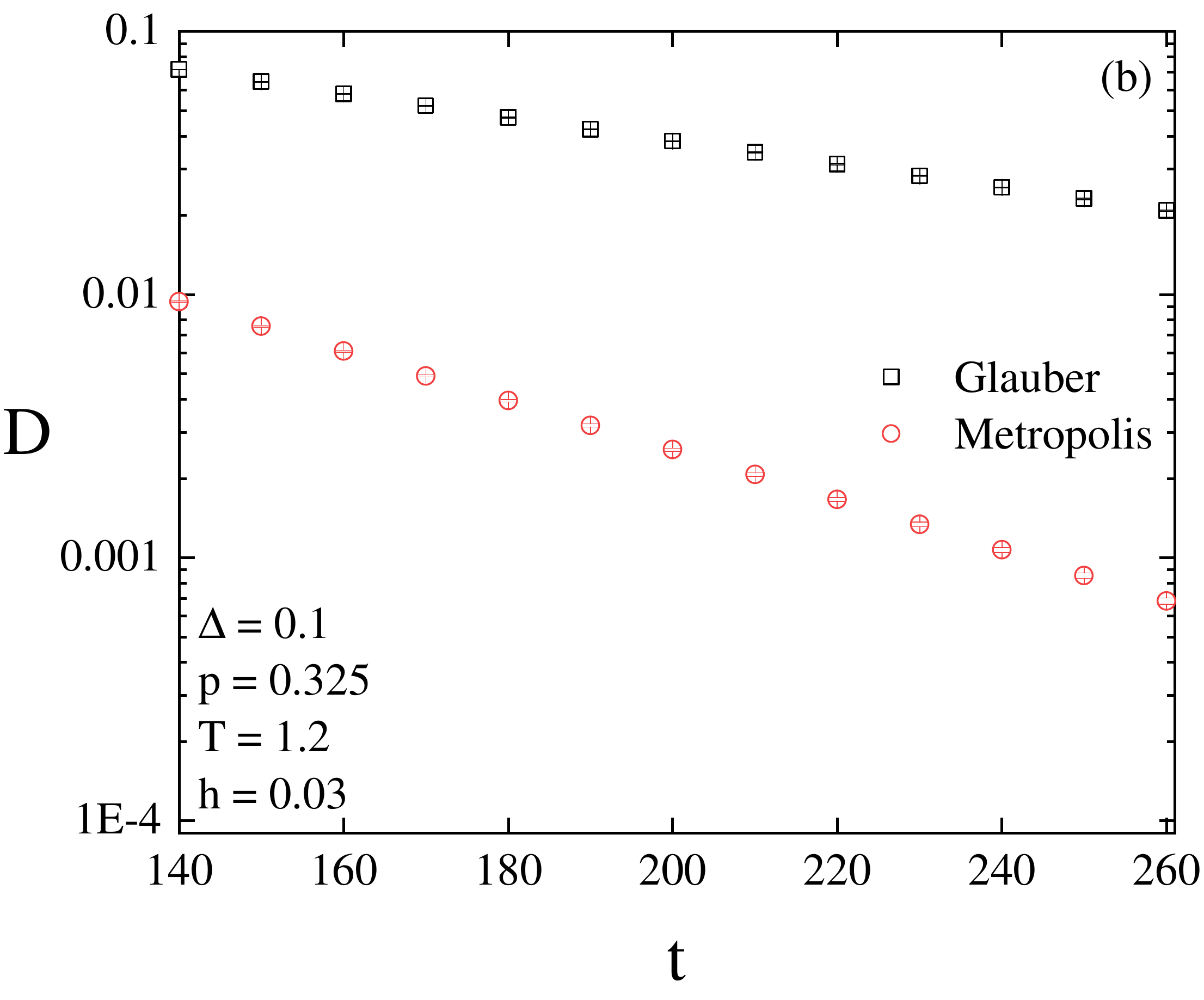}
	\caption{\label{Fig7} A comparative plot of $D(t)$ obtained via Metropolis and Glauber dynamics at $T = 1.2$, $\Delta = 0.1$, $p = 0.325$, and two values of the magnetic field: $h = 0$ (a) and $h = 0.03$ (b).}
\end{figure}

We now turn to the effect of the external magnetic field on the DS parameter. We present in Fig.~\ref{Fig4}(a) numerical data of the parameter $D$ for magnetic fields in the range $h = \{0 - 0.04\}$ and in steps of $0.01$ at fixed $T = 1.2$, $\Delta = 0$, and $p = 0.4$. As in Figs.~\ref{Fig1} and \ref{Fig3}(a), a similar behavior is observed, manifested by the solid lines that correspond to fits of the form~(\ref{eq:Dexp}) in a semi-logarithmic scale. Subsequently, Fig.~\ref{Fig4}(b) illustrates the fitting parameter $b$ obtained from the fits shown in panel (a) as a function of the applied field. Again, a linear functional behavior is recovered ($b = sh + \rm const$), and the positive slope of the straight line, $s = 0.439(23)$, is a clear indication that the magnitude of the decay constant increases linearly with $h$, again for the particular range of parameters studied.

\begin{figure}
	\centering
	\includegraphics[width=8 cm]{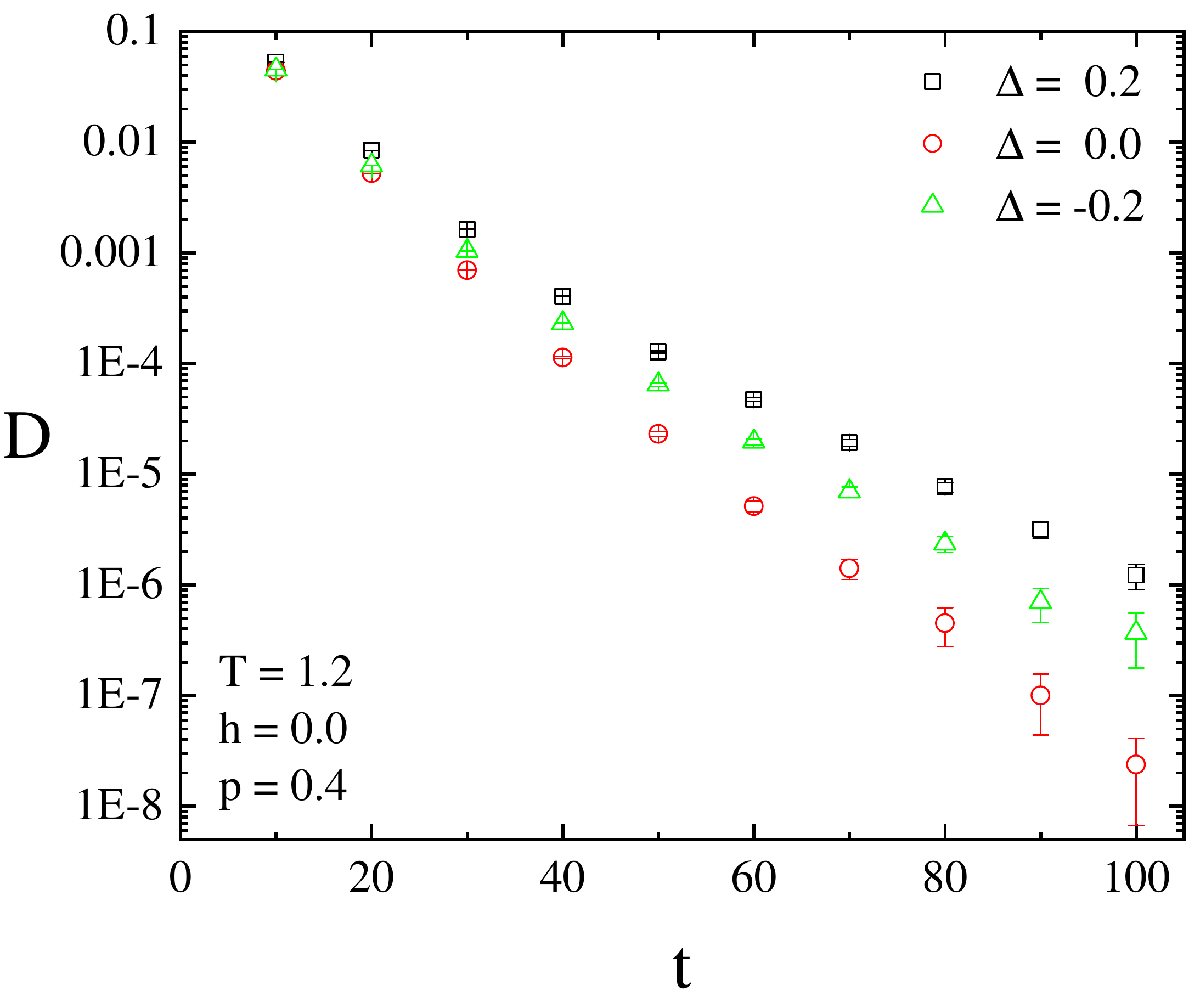}\\
	\caption{\label{Fig8} $D(t)$ curves for for the case of replica spins with $S_i = 0$ states and three values of crystal field $\Delta$, as indicated.}
\end{figure}

The last part of our analysis focuses on the effect of the amount of damage, controlled through the parameter $p$, in the DS phenomena of the system. In particular, we simulated the system~(\ref{eq:hamiltonian}) at $p = \{0.3, 0.325, 0.350, 0.375, 0.4\}$, all in the weakly-damaged regime ($p < 0.5$), and monitored the emerging transient behavior for three values of $\Delta = \{-0.2, 0, 0.2\}$ at fixed $T = 1.2$ and $h = 0$. Our analysis follows the previous discussion and is summarized in Figs.~\ref{Fig5} and \ref{Fig6}, where similar conclusions may be drawn: (i) The time dependence of $D$ for different $p$-values, see Fig.~\ref{Fig5}, follows the exponential law~(\ref{eq:Dexp}), as also shown by the solid (linear fitting) lines in the semi-logarithmic scale. As we noted before, the spins in the original lattice were initially set to 1 and, depending on amount of damage, a replica lattice was generated. As $p$ gets larger starting from zero, the distance in the spin configurations between the original and the replica lattices becomes larger. It is thus straightforward to assume that when $p$ increases, $D$ starts to get a higher value in the transient regime, as also supported by the numerical data of Fig.~\ref{Fig5}. (ii) The decay constant $b$ appears in this case as well to decrease linearly with respect to $p$, see Fig.~\ref{Fig6}.

Let us point out here that additional useful information may be obtained by deriving the multivariate functional form of the decay constant with respect to $T$, $\Delta$, and $h$. However, in order to probe the system properties within a very wide range of the simulation parameters an extensive computational task must be undertaken, that goes well beyond the context of the current work.

Before concluding, we would like to make a couple of remarks stemming from our numerical simulations and documented in Figs.~\ref{Fig7} and \ref{Fig8}: (i) We present in Fig.~\ref{Fig7} an instructive comparison on the time variation of $D$ using two spin-update rules, Metropolis and Glauber. Results are shown for two different sets of the Hamiltonian parameters implying that damage heals at a much faster rate for the case of Metropolis. This numerical observation is attributed to the fact that the spin-flip probability in the Metropolis dynamics may become unity, which on the other hand is not the case in Glauber kinetics, remaining always less than unity. (ii) We checked whether there is still an exponentially decaying average Hamming distance if one uses the zero-state spins ($S_i = 0$) in the replica lattice instead of the usual $S_i = -1$ considered throughout this work. Figure~\ref{Fig8} displays several $D(t)$ curves obtained for three different values of the crystal field $\Delta$. We observe that healing is very fast and that the system does not show an exponential behavior. A similar analysis (not shown here for brevity) was also performed for varying $h$ and $p$ values, where a similar behavior of fast healing was recorded.

\vskip 1cm

\section{Summary and outlook}
\label{sec:4}

To summarize, we have studied the transient behavior of propagation of damage in the square-lattice Blume-Capel ferromagnet. The critical behavior of this model is defined by a set of parameters, namely the strength of the single-ion anisotropy $\Delta$, the temperature $T$, and the applied magnetic field $h$. It is worth noting that for the needs of the current study we restricted ourselves to a well-defined area in the second-order phase transition regime of the phase boundary of the model, thus at a safe distance from the tricritical point. On the technical side, damage was measured via the well-known average Hamming distance $D$ and the time evolution of $D$ was monitored by extensive Monte-Carlo simulations of a large system size, $L = 1024$, using the Metropolis single spin-flip algorithm. The initial amount of damage was created by changing the spin values of a fraction $p$ of sites randomly in the replica lattice. By doing so, one generates two copies of the same system with different initial conditions depending on $p$ but with an identical random number sequence to be used in the spin dynamics. Our numerical results suggest that for the particular part of the phase diagram enclosed within $\Delta \in [-0.2, -0.2]$ and $T \in [0.9, 1.3]$ in the $\Delta - T$ plane, the average Hamming distance decays exponentially with time for all sets of the Hamiltonian parameters mentioned above but also with respect to $p$ (a scenario which is not verified if one uses the zero-state spins in the replica lattice). Moreover, the decay constant of the average Hamming distance  appears to vary linearly with respect to $T$, $\Delta$, and $h$, but also with $p$ as well, recording either an increasing ($h$-dependence) or a decreasing ($T$-, $\Delta$-, and $p$-dependence) trend. Finally, a comparative study of the propagation of damage under different dynamics rules, Metropolis and Glauber, verified the theoretical expectation that damage heals faster for the Metropolis algorithm.

As an outlook for future research, we would like to propose some possible  extensions of the present work, which also set the reported results in a a more general context: (i) It would be interesting to investigate the time scale of damage propagation in the vicinity of the percolation threshold, where the spreading/healing of damage may be described by large (perhaps diverging) times. (ii) Damage spreading under a randomly distributed anisotropy may also worth a detailed separate investigation. (iii) It would also be educative to examine the behavior of the average Hamming distance in the case where the system reaches its equilibrium configuration starting from a completely random orientation~\cite{Wang}.
(iv) It is well-known that damaging transitions fall into the
universality class of directed percolation. Grassberger showed with numerical simulations that the critical behavior at the spreading threshold $T_{\rm s}$ is consistent with directed percolation for Glauber dynamics  in the Ising model~\cite{Grassberger2}, results that have been nicely framed by the mean-field approach of Vojta~\cite{Vojta}. These works indicate that, possibly, a scaling of the form $b \sim (T_{\rm s}-T)^{\beta}$ may be expected for the decay constant upon approaching the phase boundary of the Blume-Capel model, with $ \beta$ the directed percolation critical exponent. However, as already mentioned above, dynamics in the Blume-Capel model is not as well understood in the current literature as the standard Ising ferromagnet. In order to see whether the universality class of the spreading transition of the spin-$1$ Blume-Capel model belongs to directed percolation, $T_{\rm s}$ values must be obtained to a good numerical accuracy and extensive simulations should be performed close to the phase boundary. (v) Finally, at very low temperatures (as $T \rightarrow 0$), the mean-field analysis by Vojta  for the Lyapunov exponents for damage spreading in the Ising model in the magnetized phase suggests that the decay constant $b$ should approach a constant~\cite{Vojta} and this may also be the expected for the present spin-$1$ model as well. 

\begin{acknowledgments}
A. Bhowal Acharyya acknowledges computational facilities provided by DBT, Govt. of India, for developing the code.
A part of the numerical calculations reported in this paper were performed at T\"{U}B\.{I}TAK ULAKB\.{I}M (Turkish agency), High Performance, and Grid Computing Center (TRUBA Resources). Part of this work was completed during the visit of E. Vatansever at the Research Center for Fluid and Complex Systems of Coventry University, financially supported by the Scientific and Technological Research Council of Turkey. N.~G. Fytas acknowledges support through the Visiting Scholar Program of Chemnitz University of Technology during which part of this work was completed.
\end{acknowledgments}

\vskip 1cm

\noindent {\bf Data availability:} Data will be available on request to
Muktish Acharyya.

\vskip 0.5cm

\noindent {\bf Conflict of interest:} The authors have no financial or proprietary interests in any material discussed in this article.

\vskip 0.5cm

\noindent {\bf Funding:} No funding was received particularly to support this work.

{}

\end{document}